# Abnormal Colon Polyp Image Synthesis Using Conditional Adversarial Networks for Improved Detection Performance

**Younghak Shin[1,2], Member, IEEE, Hemin Ali Qadir[2,3] and Ilangko Balasingham[1,2], Senior Member, IEEE**

[1]Department of Electronic Systems, Norwegian University of Science and Technology (NTNU), Trondheim, Norway
[2]Intervention Centre, Oslo University Hospital, Oslo Norway
[3]Department of Informatics at the University of Oslo (UiO), Oslo, Norway

Corresponding author: Younghak Shin (e-mail: shinyh0919@gmail.com).

This work was supported by the Research Council of Norway through the MELODY project under the contract number 225885/O70 and Research Council of Norway through the industrial Ph.D. project under the contract number 271542/O30.

**ABSTRACT** One of the major obstacles in automatic polyp detection during colonoscopy is the lack of labeled polyp training images. In this study, we propose a framework of conditional adversarial networks to increase the number of training samples by generating synthetic polyp images. Using a normal binary form of polyp mask which represents only the polyp position as an input conditioned image, realistic polyp image generation is a difficult task in generative adversarial networks approach. Therefore, we propose an edge filtering based combined input conditioned image. More importantly, our proposed framework generates synthetic polyp images from normal colonoscopy images which have the advantage of being relatively easy to obtain. This means realistic polyp images can be generated while maintaining the original structures of the colonoscopy image frames. The network architecture is based on the use of multiple dilated convolutions in each encoding part of our generator network to consider large receptive fields and avoid much contractions of feature map size. An image resizing with convolution for up sampling in the decoding layers is considered to prevent artifacts on generated images. We show that the generated polyp images are not only qualitatively look realistic but also help to improve polyp detection performance.

## I. INTRODUCTION

Colorectal cancer (CRC) is the second leading cancer to cause deaths for both genders [1]. CRC arises from adenomatous polyps which are growths of glandular tissue in the colonic mucosa. Most polyps are initially benign. However, some of them become malignant over time, and if left untreated, can become lethal. Therefore, the detection of early stage polyps is vital in preventing CRC. Currently, colonoscopy represents the gold standard tool for colon screening. However, colonoscopy is an operator dependent procedure and some polyps are difficult to detect even for highly trained physicians. The polyp miss-detection rate for physicians is about 25% [2]. The miss-detected polyps may lead to a late diagnosis and critical to the patient. Therefore, automatic polyp detection is important research and can be helpful to improve clinician's performance as a diagnostic supporting tool.

Recently, with the success of deep learning in many image processing and computer vision applications, convolutional neural network (CNN) based deep learning approaches have been proposed for polyp detection [3][4][5]. Yet, the detection performance is still not acceptable for use in clinical tools compared to other object detection tasks in natural image domains. The main obstacle might be the lack of available labeled colonoscopy datasets, i.e., polyp mask should be labeled by skilled clinicians, compared to natural image datasets. In addition, polyps show a large degree of variations in different scales, shapes, textures and colors. To overcome this hurdle, the concept of transfer learning schemes using natural images was introduced and evaluated for CNN based polyp detection in [6]. Generally, increasing the number of polyp training samples is highly desired in training deep networks.

In deep learning based polyp detection applications, simple image augmentation such as rotating and flipping the original images is generally used for increasing the number of training samples [6][7]. However, due to the large variation of polyps in terms of shape, scale and color, applying simple image augmentation techniques have limitations to improve system performance without changing the characteristics of the object itself and its harmony with the background.

Generative Adversarial Networks (GAN) [8] is a framework to generate artificial images by using the completive way of two networks: generator and discriminator. After a huge success of GAN, conditional GAN was proposed [9] to control the labelling of the generated images. More recently, various conditional setting based GAN frameworks were proposed in different applications such as text to image synthesis [10], style transfer [11], image super resolution [18], image to image translation [12] and segmentations [13][14].

It is known that the generator architecture is highly related to the image quality of generated images and many researches focus on the design of proper generator architectures [8][12][15]. Due to the simplicity and generalized performance, a skip connection based *U-net* architecture, which was originally proposed for medical image segmentation purpose [16], is widely used for different signal generation applications such as the image to image translation [12], voice separation [17] and image synthesis [13][14] for increasing the number of training samples [13][14].

Motivated by the conditional GAN approaches, in this study, we propose a GAN based polyp image generation framework to improve automatic polyp detection performance in colonoscopy videos. To generate realistic polyp images in which the polyp and the background are harmonious, we suggest a combined polyp conditioned images using edge filtering and polyp mask images. In addition, we propose a framework to generate synthetic polyp images from normal colonoscopy images. In this way, we can generate various abnormal polyp images while maintaining the overall content of the colonoscopy images. To the best of our knowledge, synthetic polyp image generation has not been previously addressed in the literature.

Fig. 1 shows the concept of the conditional GAN based polyp image generation. Using the proposed edge filtering based polyp conditioned input, a generator network generates realistic polyp images and a discriminator network discriminates real (target) and synthetic (output) polyp images by the adversarial training process. For design of the network architecture, we use a U-net based generator and modifies the network by applying a dilated convolution scheme [19] to avoid overly contracting the image in the encoding part of the generator. In the decoding part of the generator, we utilize an image resizing and convolution strategy instead of the transposed convolution [20][27].

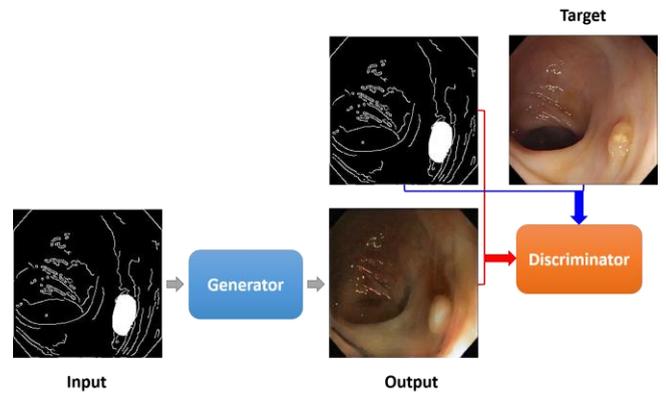

**FIGURE 1.** Conditional GANs based polyp image generation framework. Input image combines edge filtering of original image and binary polyp mask.

To evaluate the generated synthetic polyp images as an image augmentation tool, we assess automatic polyp detection performance of the generated synthetic polyp images with original polyp images. To detect polyps in colonoscopy videos, we use a recent *Faster region based CNN (Faster R-CNN)* [22] method which is a state of the art object detector in many computer vision applications [23][24].

The remainder of this paper is organized as follows. In Section II, the proposed image generation framework including network architectures and preparation of input conditioned image are introduced. In addition, we briefly explain the automatic polyp detection procedure. In Section III, experimental datasets used in this study are described. In Section IV, experimental results and discussions are presented. Finally, we conclude this study in Section V.

## II. METHODS

This section describes the conditional GAN framework and proposes network architectures for polyp generation. We introduce the suggested scheme for polyp conditioned input preparation for both training and inference. We briefly explain how we evaluate polyp detection performance of the generated polyp images using the Faster R-CNN detection scheme.

### A. Conditional Generative Adversarial Networks (GAN)

The GAN framework proposed by Goodfellow et al. [8] consists of two basic components, generator ($G$) and discriminator ($D$). The $G$ tries to fool $D$ by learning mapping from latent space to an original image space. At the same time, $D$ attempts to distinguish the real image from the generated fake image.

In the conditional GAN framework, the aim of a generator network $G$ is to learn a mapping $G : x, z \rightarrow y$ where, $x$ is an observed input, $z$ is a random noise vector and $y$ is an output generation. The loss objective of conditional GAN ($L_{cGAN}$) can be represented as follows [9]:

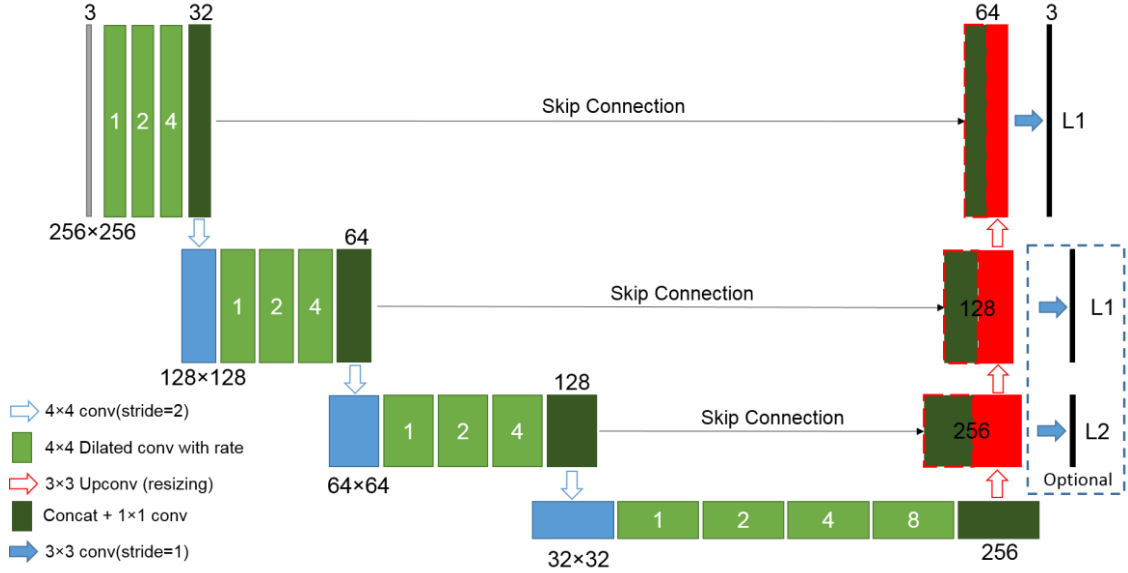

**FIGURE 2. Proposed modified U-net based generator architecture. Dilated convolution in encoding layers and resizing with convolution in decoding layers are adopted. Multi-scale L₁ and L₂ losses are optionally used.**

$$L_{cGAN}(G, D) = E_{x, y \sim p_{data}(x,y)} \left[ \log D(x, y) \right] \\ + E_{x \sim p_{data}(x), z \sim p_z(z)} \left[ \log(1 - D(G(x, z))) \right], \quad (1)$$

where $G(\cdot)$ and $D(\cdot)$ denotes the output of generator and discriminator. The $E_{x, y \sim p_{data}(x,y)}$ represents the expectation of the log-likelihood of the input and output image pair $(x, y)$ which is sampled from the underlying probability distribution of $p_{data}(x, y)$, while $p_{data}(x)$ corresponds to the distribution of input image $x$. To generate realistic images, normally L₂ [25] or L₁ [12][13] loss between generated output and original ground truth was considered in the final loss as

$$L_{L_1, L_2}(G) = E_{x \sim p_{data}(x,y), z \sim p_z(z)} \| y - G(x, z) \|_{1 \, or \, 2} \quad . \quad (2)$$

As shown in Fig. 2, we also adopt similar concept but we use more L₁ and L₂ losses in each decoding layer. We will discuss more about this intermediate losses in the last paragraph of Section II-B. The final loss function becomes

$$G^* = \arg \min_G \max_D L_{cGAN}(G, D) + \lambda L_{L_1, L_2}(G) \quad , \quad (3)$$

where $\lambda$ is a parameter to control balance between two different loss terms. In the first term, $L_{cGAN}$, $D$ tries to maximize the probability to make a correct prediction, while $G$ tries to minimize the objective competitively during the training.

### B. Network Architectures
A generator network basically follows the U-net [16] architecture which is based on the encoder-decoder network with skip connections. It is well known that the skip connections provide precise local information from each encoding layer to decoding layer. This network architecture was successfully applied to many GAN studies [12][13][14][17]. We adopt this architecture in our study but modify two main points of the generator network to improve quality of generated images. Fig. 2 shows our modified generator architecture. We use stride-2 convolution to contract the feature map size in the encoding part and skip connections for the decoding part as used in other studies [12][13].

On the other hand, we use a dilated convolution with different dilation rates in each encoding layer. The dilated convolution is a convolution with different filter size defined by the dilation rate $d$ [19]. A dilated convolution with $d = 1$ is exactly same to the normal 2-D convolution. However, if the $d$ is greater than 1, it performs convolution with $d$ holes, i.e., $d$ zeros are filled between consecutive parameter values of the convolution filters. Therefore, using a dilated convolution, we can increase the size of receptive filed while keeping the same number of parameters.

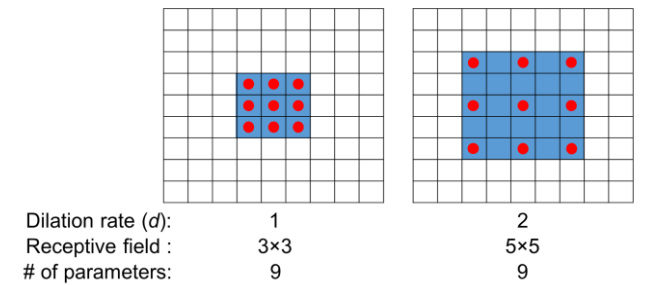

| Dilation rate ($d$): | 1 | 2 |
|---|---|---|
| Receptive field : | 3×3 | 5×5 |
| # of parameters: | 9 | 9 |

**FIGURE 3. Explanation of Dilated convolution with dilation rate 1 and 2. Dilation rate 1 is same to normal 2-D 3×3 convolution. Receptive field size is increased with dilation rate 2 while keeping the same number of parameters.**

To consider large receptive fields in CNN, normally, down sampling (i.e., pooling) is performed after the convolution layer. However, it is known that too much contraction by down sampling causes difficulty to generate detailed image in the up sampling part [19][26]. On the other hand, use of multiple dilated convolutions in the same layer has advantage in both considering large receptive fields and avoiding much contractions of the last feature map size.

To take this advantage in our polyp generation task, we use multiple dilated convolutions in each encoding part of our generator network as shown in Fig. 2. As a result, we can have less contraction of the feature map size in the last encoding layer, i.e., the feature map size of our model is 32×32, compared to the conventional U-net based architecture which has 1×1 feature map size in the last encoding part. We expect that this has the advantage of creating detailed image in the decoding part of the generator. Furthermore, due to the use of dilated convolution in our model, we can decrease the number of learnable parameters compared to the U-net based model. After applying multiple dilated convolutions in each encoding layer, we performed channel-wise concatenation for all results. We then use a 1×1 convolution to have a fixed number of channels before down sampling.

Let's focus on the up sampling part in the decoding layers of the generator network. After encoding, up sampling is crucial to generate higher resolution image which has the same size of the original image in CNN based applications such as segmentation and image synthesize. Normally the transposed convolution (also known as fractionally strided convolution) scheme is widely used for up sampling [27]. However, it is known that the transposed convolution tends to have troublesome artifacts such as checkerboard pattern [20][21]. We also observed in our experiments that the U-net based generator using the transposed convolution makes similar artifacts in the generated polyp images. Therefore, in our model, we adopt a simple resize and convolution strategy which is suggested by [20][21]. The image is first resized (by a factor of 2) for higher resolution with nearest neighbor interpolation. Then, normal 3×3 2-D convolution is performed.

Optionally, in the decoding part, we use intermediate $L_1$ and $L_2$ loss terms to train our generator network. Thus, we use a $L_2$ loss term in the first decoding layer to form initial blurred shape of generated image. At the same time, $L_1$ loss terms are used in the second and last decoding layers to encourage sharp detailed image generation. To compute intermediate loss with 64×64 and 128×128 generated images, the original ground truth image is resized to the same size of the generated images. We observe that even though the quality of generated images from this optional strategy is similar to use of the one last $L_1$ loss term, we obtain a bit smaller final training loss with the multiple loss terms. For the discriminator network, we simply utilize the widely used convolution based classification architecture suggested in [12]. For both generator and discriminator, we use an Exponential Linear Unit (ELU) activation function [28] after convolution operation.

### C. Input Conditioned Image Preparation

For training conditional GAN framework, a pair of images, i.e., input conditioned image and original ground truth image (represented by Input and Target respectively in Fig. 1), are needed. We used ground truth of polyp masks, e.g., Fig. 3-(c), which represent position of polyps in each image frame by skilled clinicians as input conditioned images.

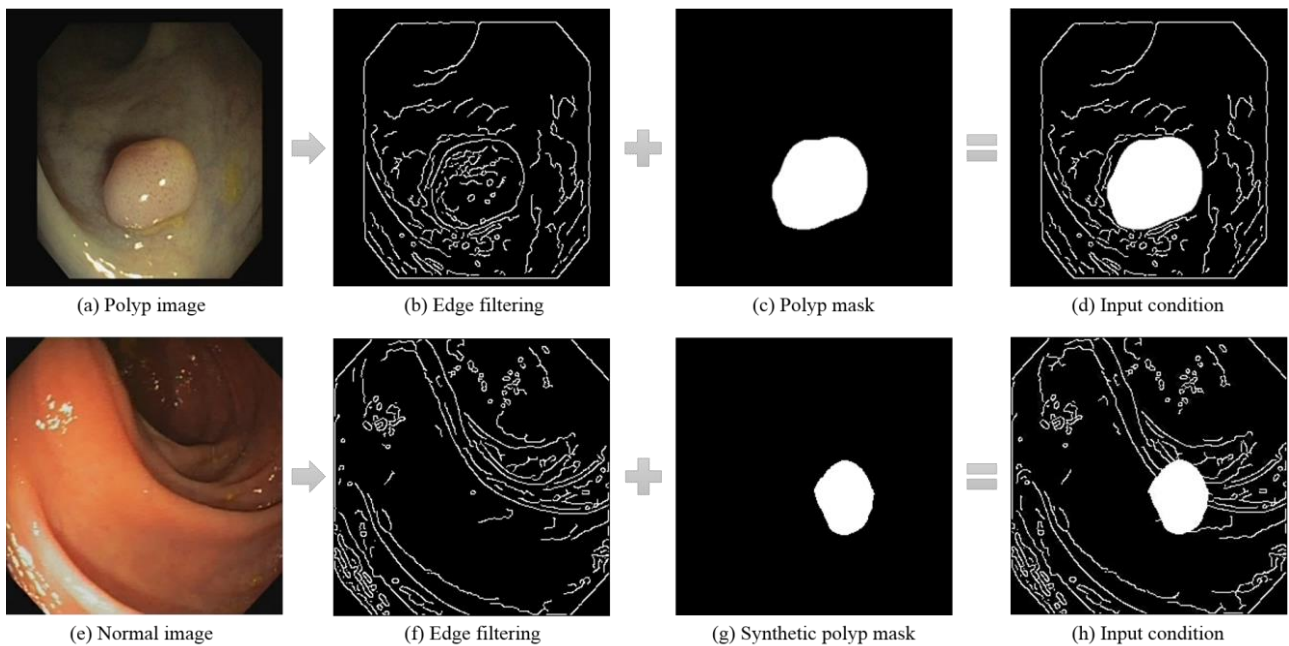

(a) Polyp image     (b) Edge filtering     (c) Polyp mask     (d) Input condition

(e) Normal image     (f) Edge filtering     (g) Synthetic polyp mask     (h) Input condition

**FIGURE 4.** Procedure of generating input conditioned image for training (a)-(d) and inference (e)-(h). First, edge filtering image is obtained from original image. Then, polyp mask is combined with the edge filtering image.

However, we found that if we only use the polyp mask, the structure of background part does not look real and the harmony of the polyp and background parts become unnatural. To overcome this issue, we suggest a combined input conditioned image as shown in top figures (a)-(d) of Fig. 4. First, apply a conventional Canny edge detector [29] to the original polyp image frame (a) to obtain contour information of colonoscopy image (b). Then, combine this edge filtered image with the polyp mask image (c) to specify the position and shape of the polyp. With this combined input image (d), we can generate more realistic polyp images which maintain the overall context of colonoscopy image frames. Image generation results from combined input and simple polyp mask are shown in Fig. 6 and Fig. 7, respectively.

### D. Normal Image to Polyp Image Generation
In inference stage, we also need input conditioned images to generate synthetic polyp images. Our final goal is to improve polyp detection performance using generated synthetic polyp images. For this, we aim to generate new unique polyp images without use of original polyp image frames. Therefore, we propose a procedure to generate input conditioned images for inference time using normal colonoscopy image frames which are relatively easy to obtain because mask labeling by skilled clinicians is not required.

Fig. 4 bottom figures (e)-(h) show the procedure to generate an inference input conditioned image. Using any normal (without polyp) colonoscopy image shown in (e), the edge filtered image (f) using the Canny edge detector is obtained. We combine a synthetic polyp mask (g) with the edge filtered image. To make new and unique shapes of polyp, we generate synthetic polyp masks using the training polyp masks by applying different combinations of image augmentation techniques such as rotation, scaling, position translation, and perspective transform with randomly selected parameters [30].

### E. Polyp Object Detector
To investigate whether the generated synthetic polyp images are effective as an augmentation tool, we evaluate the polyp detection performance. A comparison of the polyp detection performance trained by two different training datasets, i.e., the original training samples and the new training samples consisting of original samples and newly generated polyp images, is performed. For evaluation of polyp detection performance, we use a *Faster R-CNN* detection method [22] which is the state-of-the-art deep CNN based object detection algorithm [23][24].

Fig. 5 illustrates the Faster R-CNN based object detection framework using polyp images. To train the networks, polyp images and the corresponding polyp locations, i.e., rectangular shaped bounding box represented by 4 location values ($xmin$, $xmax$, $ymin$, $ymax$) are needed. Faster R-CNN method employs a region proposal network (RPN) to propose candidate object regions. The RPN works within the pre-trained deep CNN, i.e., usually the feature map of the last convolutional layer is used for the RPN [22]. Using the features extracted by the CNN and the corresponding object regions, classification and box regression layers are trained to detect polyp with corresponding polyp scores and regions. For the pre-trained deep CNN, we use a recent *Inception Resnet* [31] trained by Microsoft's (MS) COCO (Common Objects in Context) dataset [32]. This training dataset contains 112K images of 90 different common object categories such as dogs, cats, cars, *etc*. We fine-tune whole detector networks using our polyp training datasets. More detailed information about the Faster R-CNN and pre-trained network are available in [24][31].

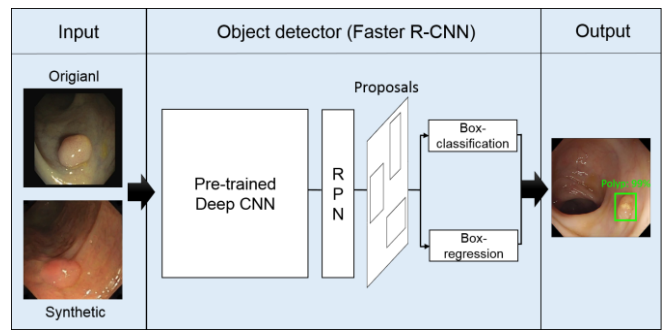

**FIGURE 5. Polyp detection framework using Faster R-CNN object detector. Pre-trained deep CNN (Inception-Resnet) is used for Faster R-CNN. Then, whole networks is fine-tuned using polyp training dataset.**

### F. Training Setup
For training of our conditional GANs, Adam optimizer [33] with 0.5 of momentum and 0.0002 of learning rate is adopted. Batch size is set to 1. In the generator network, instance normalization is used after convolution. In each encoding layers except first, 0.5 of dropout is applied after normal 2-D convolution. Before the training, the input images of $256 \times 256$ are resized to $312 \times 312$ and then randomly cropped back to $256 \times 256$ for applying random jittering [12].

For training of Faster R-CNN, we use the stochastic gradient descent (SGD) method [34] with a momentum of 0.9 with batch size of 1. In each iteration of the RPN training, 256 training samples are randomly selected from each training image where the ratio between positive ('polyp') and negative ('background') samples is 1:1. We set the learning rate equal to 1e-3. For other parameters such as non-maximum suppression (NMS) and maximum number of proposals, we use default values which were used in the original Faster R-CNN work [22].

## III. EXPERIMENTAL DATASETS
In this study, we used publicly available polyp-frame dataset, CVC-CLINIC [35], and a colonoscopy video databases, CVC-ClinicVideoDB dataset [36].

The CVC-CLINIC dataset contains 612 polyp image frames with a pixel resolution of $388 \times 284$ pixels in SD (standard definition). All images were extracted from 31 different colonoscopy videos which contain 31 unique polyps. All ground truths of polyp regions were annotated by skilled video endoscopists. This dataset is used for training our GANs to generate synthetic polyp images and training the Faster R-CNN object detector to compare polyp detection performance with the generated synthetic polyp images.

Normally, large number of training samples is preferable to train deep neural networks. Therefore, we use image augmentation techniques to increase the number of training samples and corresponding polyp masks. We apply image rotations of 90, 180, 270 degrees and horizontal/vertical flips to the original images. To create different scales of polyp images, we apply scaling augmentations with specific scaling parameters; *i.e.,* 10% and 20% of zoom-out. After all augmentations, the total number of training samples and mask is 9288. Then, we generate 9288 conditioned input image by combining edge images and polyp masks to train our GANs.

The CVC-ClinicVideoDB video dataset comprises of 18 different SD videos of different polyps. In this dataset, 10025 frames out of 11954 frames contain a polyp, and the size of the frames is 384 × 288. Each frame in the video databases comes with a binary ground truth, in which each polyp is annotated by clinical experts. Each positive video includes a unique polyp. Within each video, there is a large degree of variation with respect to scale, location and brightness. In addition, some polyp frames include artifacts such as tools for water insertion and polyp removal. We extracted 372 of normal frames (without polyp) in the videos for generating input conditioned images in inference time. Except these frames, all frames are used for testing of polyp detection performance.

## IV. RESULTS AND DISCUSSION

### A. Generated Polyp Images

Fig. 6 shows some results of the generated images from our proposed GANs. In each column, a different generated image is represented in (c) which is corresponding to each input conditioned image (b) obtained from an original normal image (a) and a synthetic polyp mask. As we can see, the generated polyp images maintain the overall structure and texture of the background from the original normal colonoscopy images. Furthermore, in the polyp parts, our trained network generates light reflections to look more realistic images.

In the generated images shown in the fourth and fifth columns of Fig. 6, the overall structures which are transformed from the normal images have changed slightly compared to the generated images in the first three columns. This is primarily due to the position of the synthetic polyp mask which is randomly placed in the input conditioned images. In this case, our trained model adaptively generates realistic polyp images by changing the structure surrounding the polyp.

In this study, to train our GANs and generate synthetic polyp images, we proposed an edge filtering based combined input conditioned images as shown in Fig. 4 (d) and (h).

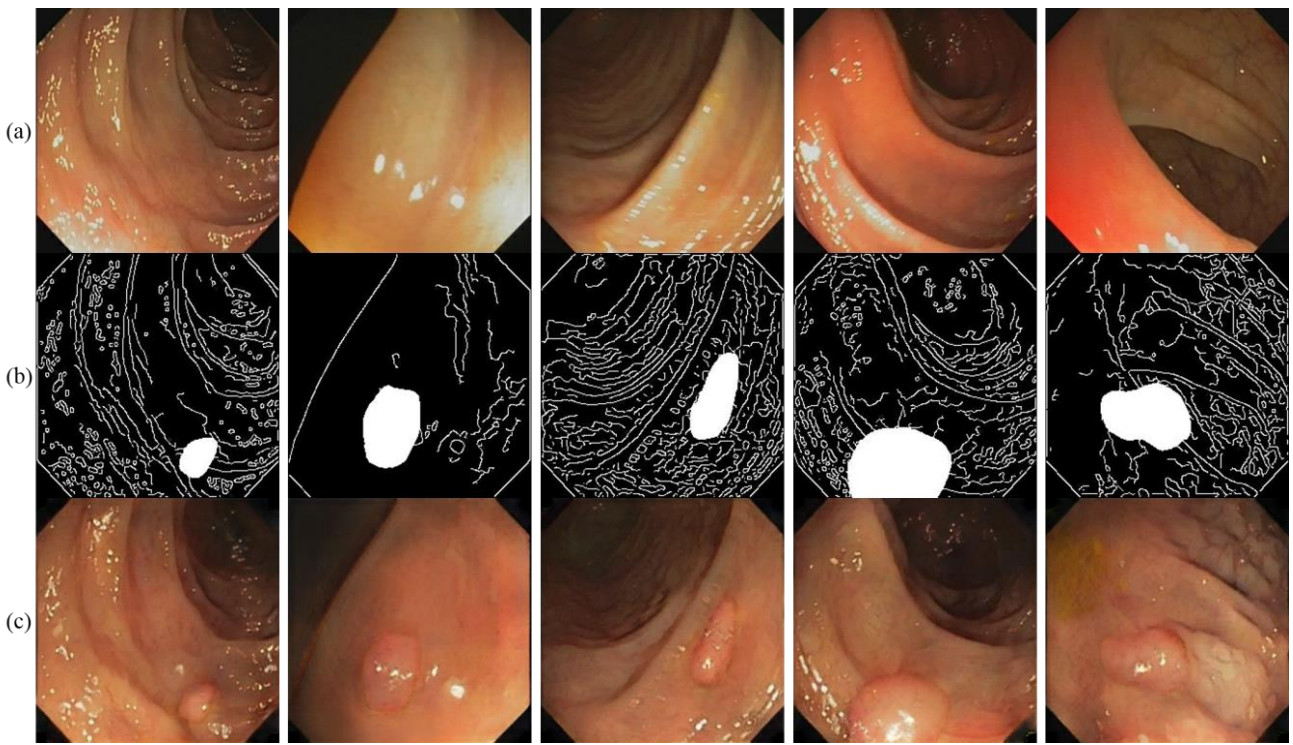

**FIGURE 6.** Results of the generated polyp images (c) from corresponding each column of the combined input image (b) obtained from the normal image (a).

To evaluate the effect of the proposed conditioned input, we used simple polyp mask images, e.g., Fig. 4 (c) and (g), for training and inference of our networks. All other training setup is exactly same to the proposed GANs.

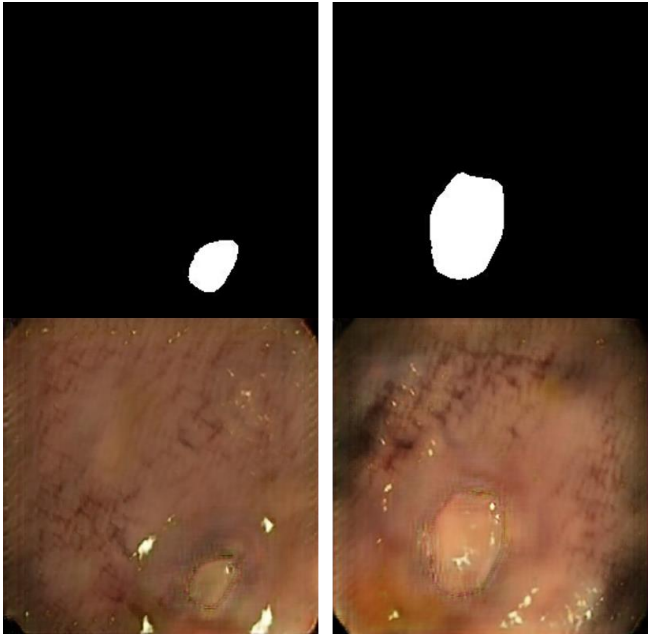

**FIGURE 7.** Results of the generated polyp images from the corresponding each column of simple polyp mask. Without information of background structures generated image quality is not successful.

Fig. 7 shows the example of two generated images. Each column shows the different generated image from the upper simple polyp mask. This polyp mask is the same one used for suggested conditioned input in the first and second column of Fig. 6. As we can see in Fig. 7, even though the network tries to generate polyp and some light reflections quite well, the background parts does not look like real colonoscopy frames compared to first and second column of Fig. 6. Therefore, in our combined input strategy, the edge information obtained from colonoscopy image frames works as an efficient guiding tool for generating overall structure of polyp images.

Fig 8 shows the comparison of the generated images from our proposed network (c) and the conventional U-net based baseline network (b). Each row represents the generated image corresponding to the same input conditioned image (a).

Based on the input combined images, both models can generate polyp images while maintaining the overall structure of the colonoscopy frames. However, in the generated images from baseline network, we can see some artifacts within the polyp parts and unclear image generation surrounding polyps. The results do not look very realistic polyp image generations compared to the generation results obtained from our network. In addition, our proposed network uses smaller number of encoding and decoding layers thanks to the dilated convolution, which results in smaller number of learnable parameters

(7494336), ca. 48%, compared to the baseline network (14304960).

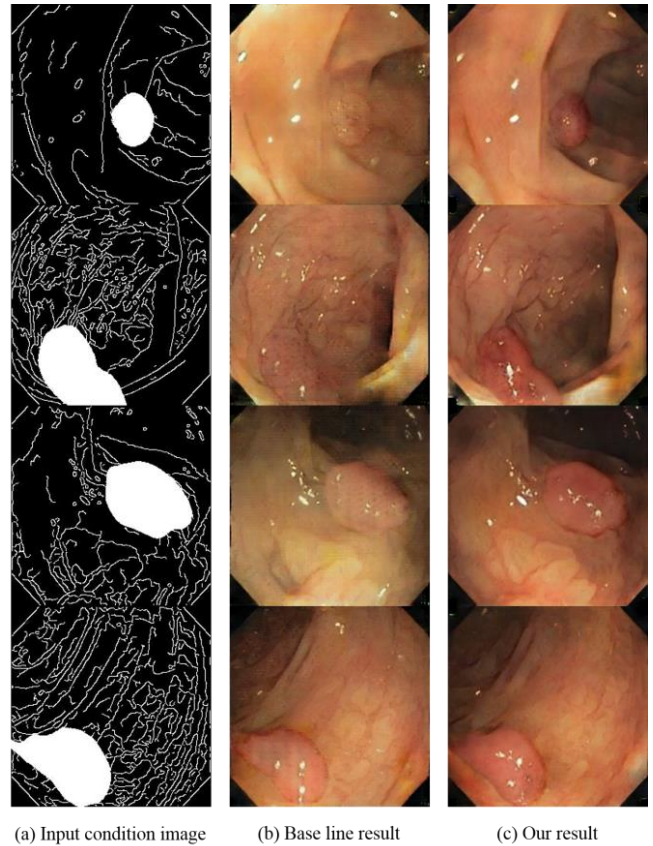

(a) Input condition image    (b) Base line result    (c) Our result

**FIGURE 8.** Comparison of generated polyp images from the baseline model (b) and our model (c). Each row shows the generated polyp images based on the given input conditioned image (a).

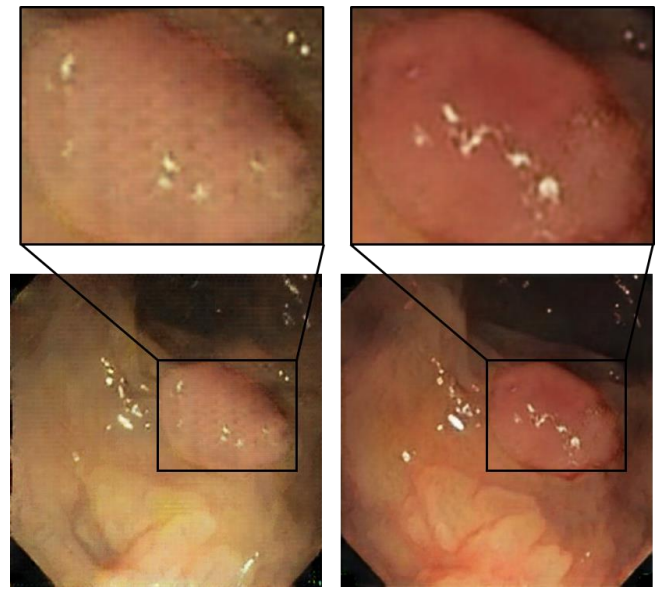

**FIGURE 9.** Example of one generated polyp image from baseline model (left) and our model (right). We observe the clear checker board artifacts in enlarged polyp area from the baseline model but not those from our model.

To see the difference clearly between the generated polyp images by both networks as shown in Fig. 9, we investigate one example of generated image in Fig. 8 (third row) by enlarging the polyp area. We observe checker board artifacts in the left figure of Fig. 9, i.e., the generated image by the baseline network. This observation is consistent with the recent literature [20][21] which report the same checker board artifacts when the transposed convolution was used for up sampling. However, our network has removed this artifact by adopting simple resize and convolution strategy as shown in the right figure of Fig. 9. We observe similar results in all generated images shown in Fig 8.

### B. Evaluation of Polyp Detection Performance
In this section, we aim to evaluate the polyp detection performance to investigate whether the generated polyp images are effective to improve polyp detection performance. For training polyp detection network (Faster R-CNN), we use the 612 original polyp images (CVC-CLINIC dataset) and the 372 generated polyp images with the corresponding polyp bounding boxes. The CVC-ClinicVideoDB (18 videos) is used for testing polyp detection performance.

To evaluate the polyp detection performance, we introduce true positive (TP), false positive (FP) and false negative (FN) where:

**TP**= detection output within the polyp ground truth.
**FP**= any detection output outside the polyp ground truth.
**FN**= polyp not detected for positive (with polyp) image.
**TN**= no detection output for negative (without polyp) image.

Note that if there is more than one detection output, only one TP is counted per polyp. Based on the above parameters, we define two performance metrics, *precision* (pre) and *recall* (rec):

$$Pre = \frac{TP}{TP+FP}, \ Rec = \frac{TP}{TP+FN} \qquad (4)$$

TABLE I
COMPARISON OF POLYP DETECTION PERFORMANCE BETWEEN ORIGINAL TRAINING SET AND COMBINED TRAINING SET BY GENERATED IMAGE.

| Training dataset | TP | FP | FN | TN | Pre (%) | Rec (%) |
|---|---|---|---|---|---|---|
| Original | 4308 | 2962 | 5717 | 1365 | 59.3 | 48 |
| Combined (Original + Generated) | 6760 | 2981 | 3265 | 962 | **69.4** | **67.4** |

Table I lists the evaluation performance of the polyp detection by comparing two training datasets, i.e., *original* (612 original images) and *combined* (612 original images + 372 generated images). The use of *combined* training dataset shows better polyp detection performance in terms of both precision and recall than the just use of *original* dataset. Specifically, after adding synthetic polyp images in training dataset, 2452 more TPs, i.e., correctly detected polyps (with just 19 more FPs, i.e., miss detected polyps), are observed and therefore, both precision and recall are improved much (10.1 and 19.4%) compared to the use of *original* image only.

As we mentioned in Section II-E, the Faster R-CNN was pre-trained by the large size natural images. However, generally, a large size training samples is preferred. Therefore, we further apply some image augmentation techniques to increase the number of training samples for both datasets. Two different image augmentation strategies are used to train the Faster R-CNN. First, we apply three rotations of 90, 180, 270 degrees and horizontal/vertical flips to the training dataset. This dataset is represented as Aug-I in Table II. Second, we apply the same three rotations and two flips to the training dataset. To increase more training samples, we applied 10% zoom-out to the original training dataset and the three rotated and two flipped dataset (Aug-II in Table II).

TABLE II
COMPARISON OF POLYP DETECTION PERFORMANCE FOR DIFFERENT AUGMENTATION STRATEGIES BETWEEN ORIGINAL TRAINING SET AND COMBINED TRAINING SET.

| Training dataset | | TP | FP | FN | TN | Pre (%) | Rec (%) |
|---|---|---|---|---|---|---|---|
| Aug-I | Original | 6113 | 2981 | 3912 | 1143 | 67.2 | 61 |
| | Combined | 7517 | 1995 | 2508 | 1013 | **79** | **75** |
| Aug-II | Original | 6011 | 1333 | 4014 | 1496 | 81.9 | 60 |
| | Combined | 6831 | 1177 | 3194 | 1399 | **85.3** | **68.1** |

Similarly in Table I, *combined* training dataset shows better polyp detection performance, i.e., precision and recall, than the *original* training dataset for both augmentation strategies. Furthermore, the use of generated images results in the increased number of TPs at the same time the decreased number of FPs compared to the just use of *original* images. For the result comparison of Aug-I and Aug-II, we observe the decrease of FPs in Aug-II compared to the Aug-I for both *original* and *combined* datasets. It might be a reason of overfitting to the training datasets since we apply zooming augmentation to the three rotated and two flipped dataset to make very large size training datasets.

Fig. 10 shows some example images of correctly detected polyps by the *combined* image dataset but not by the *original* dataset. We choose the Aug-I results since it have more TPs than Aug-II for both datasets. These polyps are missed by the trained network with *original* training dataset only. However, as shown in Fig. 10, even though the polyps look difficult to detect, they are detected by the *combined* training dataset with very high polyp detection scores, i.e., 99%. This clearly shows that our generated images actually allow more polyps to be detected. From the results of Table I, II and Figure 10, we can conclude that the generated polyp images are not only qualitatively look realistic but also help to improve polyp detection performance.

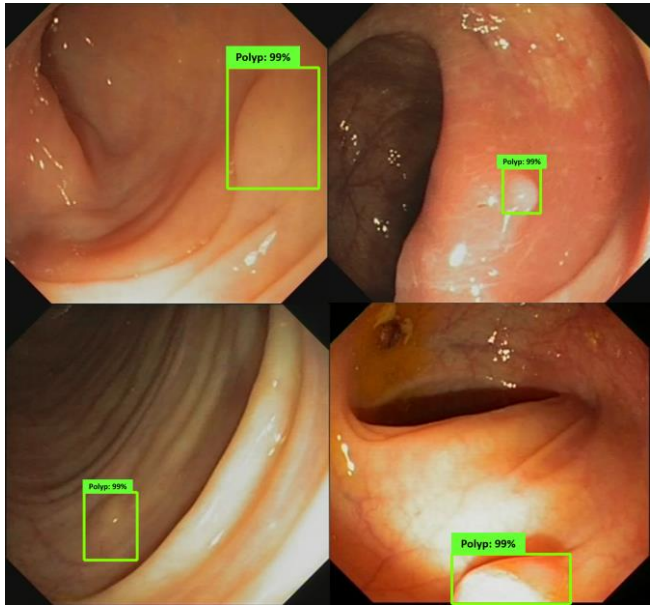

**FIGURE 10.** Example of polyp detection results by the Faster R-CNN detector. All four images include correctly detected polyps by the combined training set (Original + Generation image) but not by the Original training set.

### C. Limitations and Future Directions

Even though we successfully generate realistic polyp images using the proposed conditional GAN framework, there are some limitations. The main limitation is a generation of deterministic polyps. As we can see in Figure 6 and 8, there are not many variations in the generated polyp features in terms of color and texture. It might be because the training dataset has limited types of polyp. As we mentioned in Section III, we use 612 polyp training images. However, these images are obtained from only 31 different colonoscopy videos. More importantly, in the input conditioned image, the polyp masks labeled by clinicians only have simple binary shape information. Therefore, in the training phase, the generator is just enforced to fool the discriminator and not tries to generate variant feature types of polyp.

This issue can be solved by categorizing different types of polyps and adding a new condition in the input images. For this aim, we need to collaborate with expert clinicians for polyp categorizations. We can also try to use recent feature embedding techniques in training phase to learn a low-dimensional latent code for synthesizing diverse modes of generated images [37] and object-level mode control [38]. We think both approaches are interesting future research directions for realistic and diverse polyp generation work. However, a collection of more and variant types of polyp images should be preceded.

## V. CONCLUSION

In this work, we propose a framework to generate synthetic polyp images using a conditional GAN approach. For generation of realistic polyp images, we suggest a new generator architecture by adopting the dilated convolutions in the encoding layers and the image resizing with the convolution strategy in the decoding layers. We also suggest an edge filtering and polyp mask based combined input conditioned image to train GAN. Using this proposal we generate synthetic polyp images from various normal colonoscopy frames. Our experimental shows that the proposed GAN framework can generate more realistic polyp images than the baseline network. Furthermore, the suggested input conditioned image is helpful for preserving the overall structure of the real colonoscopy images. Finally, we demonstrate that the generated polyp images can be used as an image augmentation tools to increase the number of training samples, which helps to improve the performance of the polyp detection task.

## ACKNOWLEDGMENT

The authors would like to express sincere appreciation to Dr. Jacob Bergsland at Intervention Centre, Oslo University Hospital for his valuable comments.

## REFERENCES

[1] R. L. Siegel, K. D. Miller and A. Jemal, "Cancer statistics 2017," *CA Cancer J Clin.*, vol. 67, pp. 7-30, 2017.

[2] M. Gschwantler, S. Kriwanek, E. Langner, B. Göritzer, C. Schrutka-Kölbl, E. Brownstone, H. Feichtinger and W. Weiss, "High-grade dysplasia and invasive carcinoma in colorectal adenomas: a multivariate analysis of the impact of adenoma and patient characteristics," *Eur. J. Gastroenterol. Hepatol.*, vol. 14, no. 2, pp. 183–188, 2002.

[3] S. Park, M. Lee, and N. Kwak, "Polyp detection in colonoscopy videos using deeply-learned hierarchical features," *Seoul Nat. Univ.*, 2015.

[4] S. Park and D. Sargent, "Colonoscopic polyp detection using convolutional neural networks," *SPIE Med. Imag.*, p. 978528, 2016.

[5] J. Bernal, N. Tajkbaksh,, F. J. Sánchez, J. Matuszewski, H. Chen, L. Yu, Q. Angermann, O. Romain, B. Rustad, I. Balasingham, K. Pogorelov, S. Choi, Q. Debard, L. M. Hen, S. Speidel, D. Stoyanov, P. Brandao, H. Cordova, C. S. Montes, S. R. Gurudu, G. F. Esparrach, X. Dray, J. Liang and A. Histace, "Comparative Validation of Polyp Detection Methods in Video Colonoscopy: Results from the MICCAI 2015 Endoscopic Vision Challenge," *IEEE Trans. Med. Imaging*, vol. 36, no. 6, pp. 1231-49, 2017.

[6] N. Tajbakhsh, J. Y. Shin, S. R. Gurudu, R. T. Hurst, C. B. Kendall, M. B. Gotway and Jianming Liang, "Convolutional neural networks for medical image analysis: Full training or fine tuning?" *IEEE Trans. Med. Imag.*, vol. 35, no. 5, pp. 1299–1312, May 2016.

[7] L. Yu, H. Chen, Q. Dou, J. Qin and P. A Heng, "Integrating online and offline 3D deep learning for automated polyp detection in colonoscopy videos," *IEEE J. Biomed. Health Inform.*, vol. 21, no.1, pp.65-75, 2017.

[8] I. Goodfellow et al., "Generative adversarial nets,'' *in Proc. Adv. Neural Inf. Process. Syst.*, pp. 2672–2680, 2014.

[9] J. Gauthier. Conditional generative adversarial networks for convolutional face generation. *Technical report*, 2015.

[10] S. Reed, Z. Akata, X. Yan, L. Logeswaran, B. Schiele, and H. Lee. Generative adversarial text-to-image synthesis. *In Proceedings of The 33rd International Conference on Machine Learning*, 2016.

[11] C. Li and M. Wand. Precomputed real-time texture synthesis with markovian generative adversarial networks. *In European Conference on Computer Vision (ECCV)*, 2016.

[12] P. Isola, J. Zhu, T. Zhou, and A. A. Efros. Image-to-image translation with conditional adversarial networks. *In Conference on Computer Vision and Pattern Recognition (CVPR)*, 2017.

[13] P. Costa, A. Galdran, M. I. Meyer, M. D. Abràmoff, M. Niemeijer, A. M. Mendonca, and A. Campilho. Towards adversarial retinal image synthesis. *arXiv preprint arXiv:1701.08974*, 2017.

[14] Q Shi, X Liu and X Li, "Road detection from remote sensing images by generative adversarial networks". *IEEE access, vol. 6, pp.25486-25494, 2018.

[15] A. Radford, L. Metz, S. Chintala, "Unsupervised representation learning with deep convolutional generative adversarial networks", *In International Conference on Learning Representations (ICLR)*, 2016.

[16] O. Ronneberger, P. Fischer, T. Brox, "U-net: Convolutional networks for biomedical image segmentation", *Proc. Int. Conf. Medical Image Comput. Comput.-Assisted Intervention*, pp. 234-241, 2015.

[17] Andreas Jansson, Eric J. Humphrey, Nicola Montecchio, Rachel Bittner, Aparna Kumar, and Tillman Weyde, "Singing voice separation with deep U-Net convolutional networks," *in Proceedings of the International Society for Music Information Retrieval Conference (ISMIR)*, pp.323–332, 2017.

[18] C. Ledig, L. Theis, F. Huszár, J. Caballero, A. Aitken, A. Te-jani, J. Totz, Z. Wang, and W. Shi. Photo-realistic single image super-resolution using a generative adversarial network. *arXiv preprint arXiv:1609.04802*, 2016.

[19] F. Yu and V. Koltun. "Multi-scale context aggregation by dilated convolutions". *In International Conference on Learning Representations (ICLR)*, 2016.

[20] M.S.M. Sajjadi, B. Schölkopf and M. Hirsch, "Enhancenet: Single image super-resolution through automated texture synthesis", *In International Conference on Computer Vision (ICCV)*, 2017.

[21] A. Odena, V. Dumoulin and C. Olah, Deconvolution and checkerboard artifacts, 2016, [online] Available: http://distill.pub/2016/deconvchecke rboard/.

[22] S. Ren, K. He, R. Girshick, and J. Sun, "R-CNN: Towards real-time object detection with region proposal networks," *in Advances in Neural Information Processing Systems*, Montreal, QC, pp. 91–99, 2015.

[23] L. Zhang, L. Lin, X. Liang, and K. He. "Is faster r-cnn doing well for pedestrian detection?," *In European Conference on Computer Vision (ECCV)*, pp. 443-457, 2016.

[24] J. Huang, V. Rathod, C. Sun, M. Zhu, A. Korattikara, A. Fathi, I. Fischer, Z. Wojna, Y. Song, S. Guadarrama and K. Murphy, "Speed/accuracy trade-offs for modern convolutional object detectors", *in Proc. IEEE Conf. on Computer Vision and Pattern Recognition (CVPR)*, 2017.

[25] D. Pathak, P. Krahenbuhl, J. Donahue, T. Darrell, and A. A. Efros. "Context encoders: Feature learning by inpainting". *In Conference on Computer Vision and Pattern Recognition (CVPR)*, 2016.

[26] L.-C. Chen, G. Papandreou, I. Kokkinos, K. Murphy, and A. L. Yuille. "DeepLab: Semantic Image Segmentation with Deep Convolutional Nets, Atrous Convolution, and Fully Connected CRFs". *arXiv preprint arXiv:1606.00915*, 2017.

[27] M. D. Zeiler and R. Fergus. "Visualizing and understanding convolutional neural networks". *In European Conference on Computer Vision (ECCV)*, 2014.

[28] D. Clevert, T. Unterthiner and S. Hochreiter, "Fast and accurate deep network learning by exponential linear units (ELUs)", *Proc. Int. Conf. Learn. Represent.*, pp. 1-14, 2016.

[29] J. Canny, "A Computational Approach to Edge Detection," *IEEE Trans. Pattern Anal. Mach. Intell.*,vol.8, No. 6, pp. 679-698, 1986.

[30] A. Jung [online] Available: https://github.com/aleju/imgaug

[31] C. Szegedy, S. Ioffe, and V. Vanhoucke, "Inception-v4, inception-resnet and the impact of residual connections on learning", *arXiv:1602.07261*, 2016.

[32] T.-Y. Lin, M. Maire, S. Belongie, J. Hays, P. Perona, D. Ramanan, P. Dollar, and C. L. Zitnick. "Microsoft COCO: Common objects in context", *In European Conference on Computer Vision (ECCV)*, 2014.

[33] D. Kingma and J. Ba. "Adam: A method for stochastic optimization". *In International Conference on Learning Representations (ICLR)*, 2015.

[34] Alex Krizhevsky, Ilya Sutskever, and Geoffrey E Hinton. "Imagenet classification with deep convolutional neural networks", *In Neural Information Processing Systems (NIPS)*, 2012.

[35] J. Bernal, J. Snchez, G. F.-Esparrach, D. Gil, C. Rodriguez and F. Vilario, "Wm-dova maps for accurate polyp highlighting in colonoscopy: Validation vs. saliency maps from physicians," *Comput. Med. Imag. Graph.*, vol. 43, pp. 99–111, 2015.

[36] Q. Angermann, J. Bernal, C. Sánchez-Montes, M. Hammami, G. Fernández-Esparrach, X. Dray, O. Romain, F. J. Sánchez and A. Histace, "Towards Real-Time Polyp Detection in Colonoscopy Videos: Adapting Still Frame-Based Methodologies for Video Sequences Analysis" *In Computer Assisted and Robotic Endoscopy and Clinical Image-Based Procedures, Springer, Cham.*, pp. 29-41, 2017.

[37] J.-Y. Zhu, R. Zhang, D. Pathak, T. Darrell, A. A. Efros, O. Wang, and E. Shechtman. "Toward multimodal image-to-image translation," *In Advances in Neural Information Processing Systems (NIPS)*, pp. 465–476, 2017.

[38] T.-C. Wang, M.-Y. Liu, J.-Y. Zhu, A Tao, J. Kautz, B. Catanzaro, "High-resolution image synthesis and semantic manipulation with conditional GANs", *Proc. IEEE Conf. Comput. Vis. Pattern Recognit. (CVPR)*, pp. 1-13, 2018.